\documentclass[aps,prl,twocolumn,groupedaddress]{revtex4-1}
\pdfoutput=1

\usepackage{amsmath}
\usepackage{amssymb}
\usepackage{lastpage} 
\usepackage{multirow}
\usepackage{graphicx}
\usepackage{float}
\usepackage{midfloat}

\begin{document}

\title{Mechanical control of a microrod-resonator optical frequency comb}

\begin{abstract}
Robust control and stabilization of optical frequency combs enables an extraordinary range of scientific and technological applications, including frequency metrology at extreme levels of precision \cite{Rosenband2008,Jiang2011}, novel spectroscopy of quantum gases \cite{Ni2008a} and of molecules from visible wavelengths to the far infrared \cite{Diddams2007}, searches for exoplanets \cite{Steinmetz2008,Li2008,Ycas2012}, and photonic waveform synthesis \cite{Jiang2007,Fortier2011}.  Here we report on the stabilization of a microresonator-based optical comb (microcomb) by way of mechanical actuation.  This represents an important step in the development of microcomb technology, which offers a pathway toward fully-integrated comb systems.  Residual fluctuations of our 32.6 GHz microcomb line spacing reach a record stability level of $5\times10^{-15}$ for 1 s averaging, thereby highlighting the potential of microcombs to support modern optical frequency standards.  Furthermore, measurements of the line spacing with respect to an independent frequency reference reveal the effective stabilization of different spectral slices of the comb with a $<$0.5 mHz variation among 140 comb lines spanning 4.5 THz.  These experiments were performed with newly-developed microrod resonators, which were fabricated using a CO$_2$-laser-machining technique.
\end{abstract}

\author{Scott B. Papp}
\email{scott.papp@nist.gov}
\author{Pascal Del'Haye}
\author{Scott A. Diddams}
\affiliation{National Institute of Standards and Technology, Boulder, Colorado 80305, USA}


\date{\today}
\maketitle

Femtosecond-laser optical frequency combs have revolutionized frequency metrology and precision timekeeping by providing a dense set of absolute reference lines spanning more than an octave.  These sources exhibit sub-femtosecond timing jitter corresponding to, for example, an ultralow phase noise of $<100$ $\mu$rad on the 10~GHz harmonic of the repetition frequency (line spacing) \cite{Fortier2011,Diddams2004}.  Achieving this remarkable performance depends jointly on low intrinsic comb noise and on frequency control of the comb to duplicate the high stability of fixed optical references across the entire output spectrum.

Recently, a new class of frequency combs has emerged based on monolithic microresonators \cite{Kippenberg2011}, henceforth denoted microcombs.  These devices have the potential to significantly reduce the bulk, cost, and complexity of conventional laser combs.  Such factors stand in the way of next-generation applications that will require high-performance optical clocks for experiments outside the lab, or even in space \cite{Schiller2009}.  Here the comb generation relies on parametric conversion provided simply by third-order nonlinear optical effects and is enabled by advances in the quality factor $Q$ and the mode volume of microresonators.  These devices require only a single continuous-wave laser source, but the achievable frequency span of the comb depends on low dispersion, making material properties critical.  To date microcombs have been explored with a number of microresonator technologies, including microtoroids \cite{DelHaye2007}, crystalline resonators \cite{Savchenkov2008,Grudinin2009,Chembo2010}, microrings \cite{Levy2010,Razzari2010}, fiber cavities \cite{Braje2009}, machined disks \cite{Papp2011}, and wedge resonators \cite{Lee2011a}.  Unique comb spectra have been demonstrated featuring octave spans \cite{DelHaye2011,Okawachi2011} and a wide range of line spacings \cite{Li2012}.  And some aspects of the microcomb frequency-domain behavior have been explained \cite{Savchenkov2008,Savchenkov2008a,DelHaye2008,Papp2011}.
  
\begin{figure}[b] \centering \vspace{-12pt}
\includegraphics[width=0.47\textwidth]{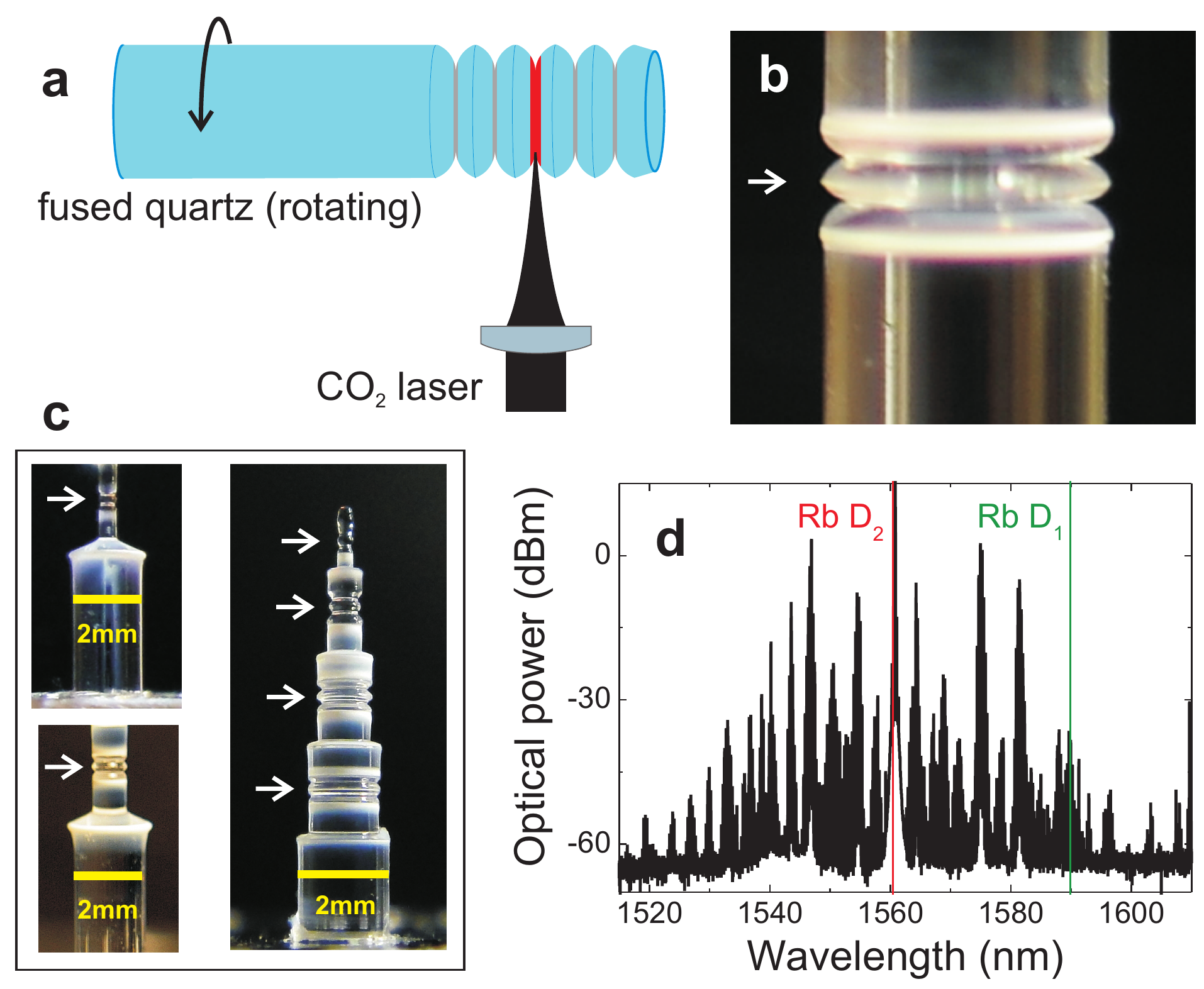}
\caption{Fabrication of fused-quartz microrod resonators by use of CO$_2$ laser machining. (a)  A rotating fused-quartz rod is illuminated with a focussed CO$_2$ laser that selectively removes material.  By applying the laser at different positions along the rod's axis, a microresonator is produced. (b)  Image of a microrod with $Q=5\times10^8$, 2 mm diameter, and $\sim$ 100 $\mu$m thickness. (c)  Fabrication of resonators with variable diameter.  Starting at top left and counting clockwise, the resonator diameters are 0.58 mm, 0.36 mm, 0.71 mm, 1.2 mm, 1.5 mm, and 1.0 mm. (d)  Optical spectrum from the device in (b), which has a modulated envelope characteristic of parametric combs \cite{Chembo2010,Chembo2010a,Papp2011,Matsko2012}.  The span of this comb is sufficient to access the D1 and D2 transitions of atomic Rb, following second-harmonic generation.
\label{fig1} \vspace{-0pt}}
\end{figure} 

Microcombs present an interesting challenge for frequency stabilization, as first pointed out by Del'Haye et al. in Ref. \cite{DelHaye2008}.  Specifically, the center frequency of a microcomb spectrum is matched to the pump laser, and the line spacing must be controlled by changing the resonator's physical properties.  Future metrology applications of microcombs will require stabilization of the line spacing with respect to fixed optical and microwave frequency standards.  Hence the key factors for stabilization are a line spacing in the measurable 10's to 100~GHz range, low intrinsic fluctuations, and the capability for fast modulation.  Additionally, a threshold power for comb generation in the milliWatt range, and the potential for integration with chip-based photonic circuits would enable portable applications.  

Here we report a new microcomb platform for achieving these goals.  We have developed a CO$_2$-laser-machining technique that yields microrod optical resonators with a $Q\gtrsim5\times10^8$, a user-defined diameter, and a small effective mode area.  The resonant optical frequencies of these devices can be rapidly controlled by using mechanical forces that alter the resonator's shape.  With such microresonators, we create a comb spectrum with 32.6-GHz-spaced lines spanning from 1510 nm to 1620~nm.  Our work introduces wideband mechanical control of the microcomb line spacing and its stabilization with respect to microwave standards.  We have improved by more than a factor of 200 the residual line-spacing fluctuations beyond all previous microcomb work.  And, for the first time, we demonstrate the potential of microcombs to support optical frequency references that feature fractional stability in the 10$^{-15}$ range.

\begin{figure}[t] \centering
\includegraphics[width=0.45\textwidth]{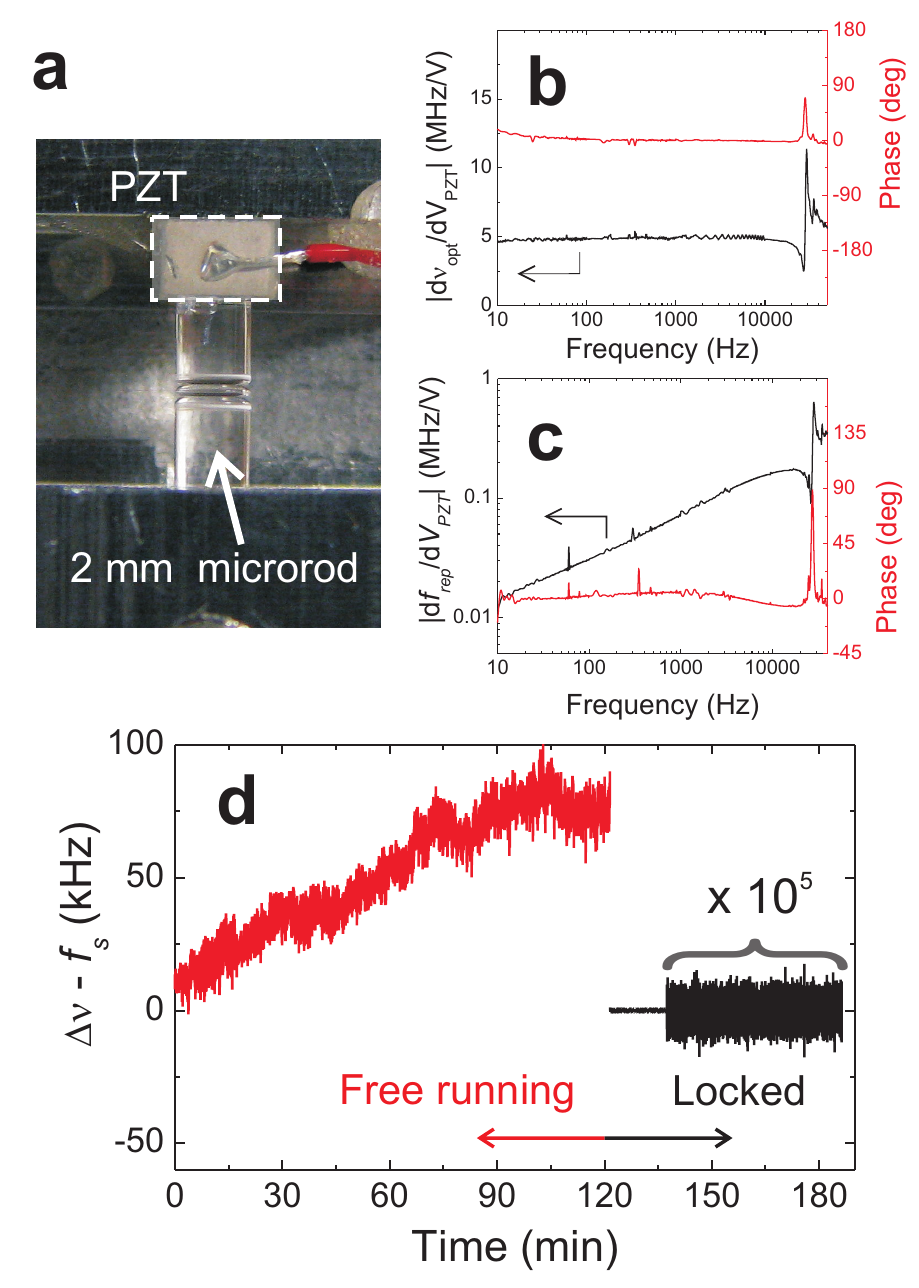}
\caption{Mechanical control mechanism for microresonator combs. (a)  Control apparatus.  A PZT compresses the rod containing our microrod resonator to adjust its mode frequencies. (b, c)  Response of optical resonance (b) and microcomb line spacing (c) with PZT drive frequency.  (d)  A three-hour record of the microcomb line spacing ($\Delta\nu$) under free running and stabilized conditions.  The line spacing set point $f_s$ is 32.5671 GHz.  A portion of the stabilized record is vertically scaled by $10^5$ to make the level of fluctuations visible.
\label{fig2} \vspace{-12pt}}
\end{figure} 

To create microrod resonators for comb generation, we use a CO$_2$ laser to simultaneously shape and polish a fused-quartz rod (Fig \ref{fig1}).  At 10.6 $\mu$m fused quartz is highly absorptive, such that melting and evaporation of the material is easily accomplished with a focussed $<5$~W CO$_2$ laser beam.  Moreover, the thermal conductivity of fused quartz is low, which enables localized heating to beyond the melting point of $\approx1600$ $^{\circ}$C.  Figure \ref{fig1}a illustrates the basic procedure for resonator fabrication.  A 2 mm diameter fused-quartz rod is rotated in a ball-bearing spindle and the CO$_2$ laser is directed normal to the axis of the rod.  The basic shape of a spheroidal resonator is created by iteratively applying 5 s laser pulses at locations laterally separated by 0.3 mm.  This process also limits re-deposition of material on the resonator surface.  At constant CO$_2$ power, the machining self terminates when the volume of fused quartz that reaches the evaporation temperature is removed.  The resonator fabrication procedure we developed has several unique features, including a $<1$ minute run time, built-in polishing of the resonator surface to support ultrahigh $Q$, and a resonator yield of nearly 100 \% with $Q$ in the 1--6 $\times10^8$ range.  A video demonstration is included in the Supplementary Information.

We discovered that self termination of the CO$_2$ process enables control over the resonator diameter with $\pm$10~$\mu$m precision.  Before machining a resonator, we can arbitrarily reduce the quartz rod diameter by positioning the CO$_2$ laser beam slightly above (or below) the rod and repeatedly moving it back and forth along the rod's axis.  All the quartz material subjected to sufficiently high laser power is removed.  Moreover, this procedure results in a smooth surface with respect to the fixed position of the laser beam.  Figure \ref{fig1}c shows fused-quartz rods that were turned down in this manner, and we have fabricated microrods that produce combs with line spacing ranging from 33 GHz to 150 GHz.  A video of the diameter reduction process is also available in the Supplementary Information.  The image at right in Fig. \ref{fig1}c demonstrates our capability to fabricate microrods of varying diameter on a single fused-quartz sample.  This feature will enable future experiments that require precise control of resonator free spectral range, such as accessing narrowband Brillouin gain \cite{Li2012} or matching the line spacing of microcombs to the ground-state hyperfine transition frequencies of atoms.

To generate microcomb spectra (Fig. \ref{fig1}d), we pump a microrod with light coupled via a tapered optical fiber \cite{Cai2000,Spillane2003}.  The pump laser, a tunable semiconductor laser operating near 1560 nm, is amplified in erbium fiber and then spectrally filtered to remove ASE noise; 280 mW of light is available at the input to the tapered fiber.  The microrod is passively locked to the pump laser via thermal bistability \cite{Carmon2004}.  This allows us to stabilize the microcomb center to an auxiliary laser, which in turn is frequency doubled and referenced to a rubidium D2 transition at 780 nm \cite{Ye1996}.  The Rb atoms provide an absolute fractional stability of $\sim10^{-11}$ at 1 s, but the 1 s residual noise of $<10^{-17}$ between the microcomb pump and the auxiliary laser indicates that much more stable references can be employed in the future.  The spectrum of our comb, which spans $\sim100$ nm, reaches the corresponding wavelength (1590 nm) of Rb D1 lines at 795 nm.  This opens the possibility for all-optical stabilization of the comb center and mode spacing using Rb transitions; a future paper will explore this idea.  In this Letter, we focus on characterization and stabilization of the 32.6 GHz line spacing, which is measured by way of direct photodetection.  After conversion to baseband (described below), the line-spacing signal is analyzed with respect to ultralow phase- and frequency-noise hydrogen-maser oscillators, which feature an $\approx10^{-13}$ at 1 s fractional frequency stability.  An important measure of comb performance is the intrinsic frequency jitter of the line spacing.  Figure \ref{fig2}d shows a two-hour record of the free running microcomb line spacing.  The 1~s Allan deviation for 100~s increments of this data, taken under typical laboratory conditions, ranges from $2\times10^{-8}$ to $10^{-7}$.

\begin{figure}[t] \centering
\includegraphics[width=0.45\textwidth]{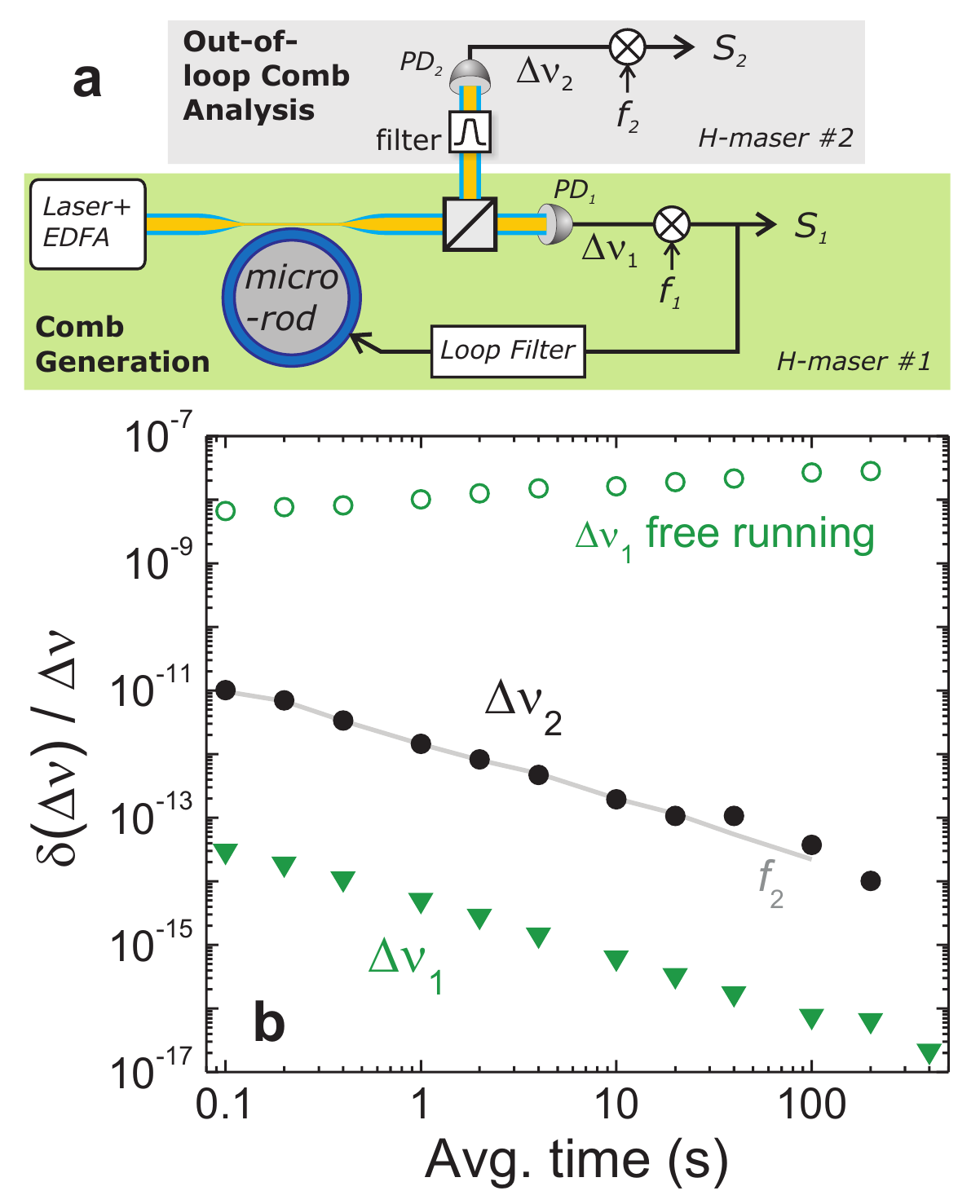}
\caption{Microcomb line spacing stability. (a)  Schematic of our system with independent paths for microcomb generation and line spacing stabilization (green box), and ``out-of-loop'' analysis (gray box).  The entire optical path is in fiber, including a programmable optical filter used to study the line spacing stability for portions of the comb (Fig. \ref{fig4}).  Frequency references $f_{1,2}$ are used for baseband conversion, and the signals $S_{1,2}$ are measured.  (b)  Line-spacing Allan deviation versus averaging time for: (triangles) stabilized residual, (points) stabilized absolute, and (open circles) free running.  The gray line shows the Allen deviation of frequency reference $f_2$.
\label{fig3}\vspace{-18pt}}
\end{figure} 

Here we introduce a mechanism for control of the comb's line-spacing noise via a mechanical force applied along the axis of the fused-quartz rod.  Mechanical control offers significant advantages including low-power operation, simple integration with bulk resonators, and response potentially much faster than resonator thermal conduction.  An image of our setup for line spacing control is shown in Fig. \ref{fig2}a.  A piezoelectric (PZT) element is used to compress the fused-quartz rod, resulting in axial expansion and tuning of the resonator's mode structure.  In Fig. \ref{fig2}b and c, we characterize the magnitude and phase modulation response of a resonator mode and the line spacing of our comb, respectively.  For a pump power well below the thermal bistability point, we monitor the resonance frequency of a mode as the PZT voltage is varied; see Fig. \ref{fig2}b.  The PZT adjusts the mode frequency by 5 MHz/V below a mechanical resonance of the system at 25 kHz.  This response is less than what is expected ($P_{PZT}\nu /E\times 2 \rm{mm}$), given the Young's modulus $E$ and Poisson ratio $\nu$ for fused quartz, and the $\sim1$ MPa/V PZT stroke.  The discrepancy is likely explained by a poor mechanical connection.  The line spacing of the comb also tunes with PZT voltage up to 25 kHz; however, the resonator thermal locking mechanism reduces the magnitude response at low frequency.  A near-zero phase delay between the modulation and the PZT-induced response indicates the passive nature of the thermal lock, and it satisfies a basic requirement for providing useful feedback.  The PZT enables stabilization of the line spacing, which is evident starting at 120 min in Fig. \ref{fig2}d. Compared to the free-running case, its drift has been reduced by a factor of $\sim10^6$.

We analyze the line spacing in detail to understand a microcomb's potential for replicating in each comb line the stability of state-of-the-art frequency references.  Figure \ref{fig3}a shows the important elements of our apparatus.  Following generation, the microcomb spectrum is delivered to two systems for independent stabilization and analysis.  In both these paths the 32.6 GHz comb line spacings ($\Delta\nu_1$ and $\Delta\nu_2$) are photodetected, amplified, and converted to the baseband signals $S_1$ and $S_2$.  Importantly, $S_{1,2}$ carry the fluctuations of both the line spacing and the microwave references ($f_1$ and $f_2$), which are locked to independent maser signals.  The Allan deviation and phase-noise spectra of signals $S_{1,2}$ are recorded separately by use of a commercial phase noise analyzer, which is referenced to maser 1 (2) for residual (absolute) measurements.  By initiating a phase-locked loop using $S_1$ and the PZT, we stabilize $\Delta\nu_1$ with respect to maser 1.  At an averaging time of 1 s, the $5\times10^{-15}$ residual fluctuations of $\Delta\nu_1$ (green triangles in Fig. \ref{fig3}b) are far below the stability of maser 1.  This signifies that the microcomb closely follows the reference frequency $f_1$ and attains its stability.  Furthermore, our analysis system tests the microcomb's ability to characterize independent microwave frequencies such as $f_2$.  In Fig. \ref{fig3}b the solid line shows the Allan deviation of $f_2$ from a separate measurement, and the filled points show the combined fluctuations of $f_2$ and $\Delta\nu_2$.  These data confirm the expectation from our residual measurements that the absolute stability of $\Delta\nu_2$ is significantly better than $1.5\times10^{-12}$ at 1 s.  The consistent $1/\rm{time}$ averaging behavior observed in both our residual and absolute measurements is evidence of the phase-locked stabilization.  In contrast, the open circles in Fig. \ref{fig3}b show the free-running line-spacing drift that increases with time. 

\begin{figure}[t] \centering
\includegraphics[width=0.45\textwidth]{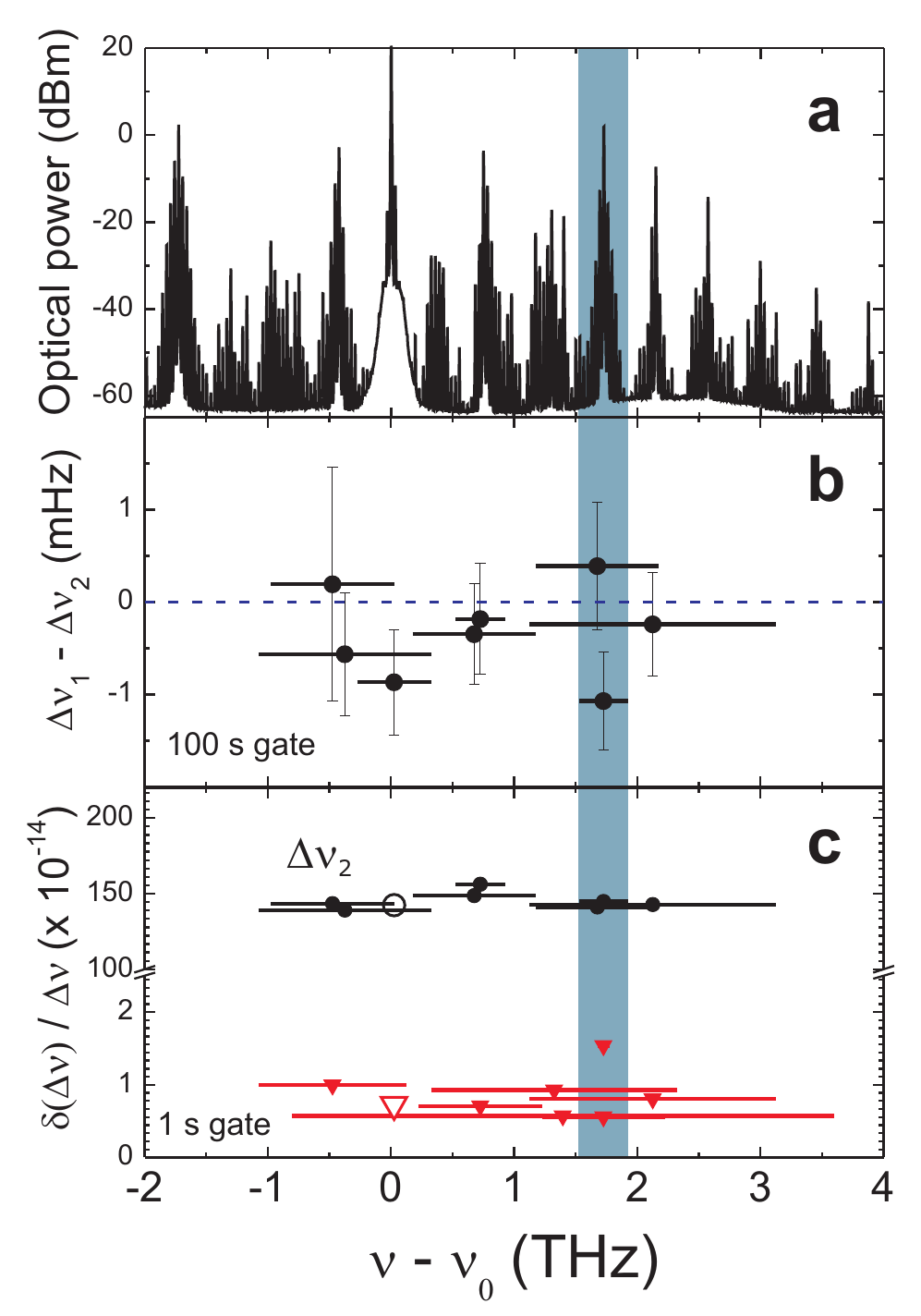}
\caption{Line-spacing equidistance and stability for different spectral slices of the comb.  (a)  Microcomb optical spectrum about the pump laser frequency $\nu_0$ prior to filtering.  (b)  Measurements of the frequency difference ($\Delta\nu_1-\Delta\nu_2$) between the whole comb and a spectral slice.  The shaded region indicates the frequency range for one measurement.  (c) For different spectral slices, the points show the 1 s Allan deviation of $\Delta\nu_2$, and the triangles characterize residual fluctuations between $\Delta\nu_1$ and $\Delta\nu_2$.  For the open circle and open triangle data points, only a 0.3 THz range about $\nu_0$ is blocked.
\label{fig4}\vspace{-18pt}}
\end{figure} 

The $S_1$ signal used for line-spacing stabilization is a composite of all the comb lines, and its largest contributions naturally come from the most intense pairs.  Hence, an uneven distribution of comb optical power, along with the complicated nonlinear comb generation process, opens the possibility of degraded line-spacing stabilization for different spectral slices of the comb.  To quantify these effects, we probe the line-spacing frequency and its stability with our comb analysis system.  By use of the 1535 nm to 1565 nm (C-band) programmable optical filter with 10 GHz resolution shown in Fig. \ref{fig3}a, we obtain an arbitrary selection of comb lines.  Figure \ref{fig4}b shows measurements of the difference in line spacing ($\Delta\nu_1-\Delta\nu_2$) between the entire comb and various portions of it.  Here the horizontal bars indicate the range of optical frequencies present in the filtered $\Delta\nu_2$ signal, and $\Delta\nu_1$ is determined by the set point of our phase-locked loop.  (The residual offset between maser 1 and locked $S_1$ is $<1$ $\mu$Hz.)  For reference, the shaded area indicates the comb lines studied in a single measurement.  The weighted mean of all data is -0.4 mHz (on the 32.6 GHz line spacing) from the anticipated null, which is consistent with their uncertainties and with our knowledge of the offset between the maser-referenced $f_1$ and $f_2$ signals.  Moreover, a fit of the slope in Fig. \ref{fig4}b demonstrates that the line spacing does not change by more than the $5\times10^{-15}$ standard error over a 4.5 THz span of the comb.  

The line-spacing stability of the spectral slices also characterizes the PZT stabilization.  Figure \ref{fig4}c shows the 1 s Allan deviation associated with each 400 s long frequency difference measurement.  The stability of $\Delta\nu_2$ throughout the C-band portion of the comb is $1.5\times10^{-12}$, a value dominated by frequency reference $f_2$.  It appears that the mechanisms responsible for line-spacing noise act similarly to different components of the comb, and our PZT control can effectively counter them.  To understand the residual stability of $\Delta\nu_2$ that is possible apart from the noise of $f_2$, we reconfigure our system to use $f_1$ for baseband conversion of both $\Delta\nu_1$ and the optically-filtered $\Delta\nu_2$.  In this case, common $f_1$ noise contributions are suppressed when the $S_{1,2}$ signals are presented to our noise analyzer.  What remains is: uncontrolled jitter between the spectral slices and the whole comb, and the noise associated with the independent optical and electrical measurement paths.  The level of these residual fluctuations is mostly below $10^{-14}$ at 1 s; see the red triangles in Fig. \ref{fig4}c.  This demonstrates that future microcomb experiments could take advantage of frequency references even more stable than a maser.

\begin{figure}[t] \centering
\includegraphics[width=0.45\textwidth]{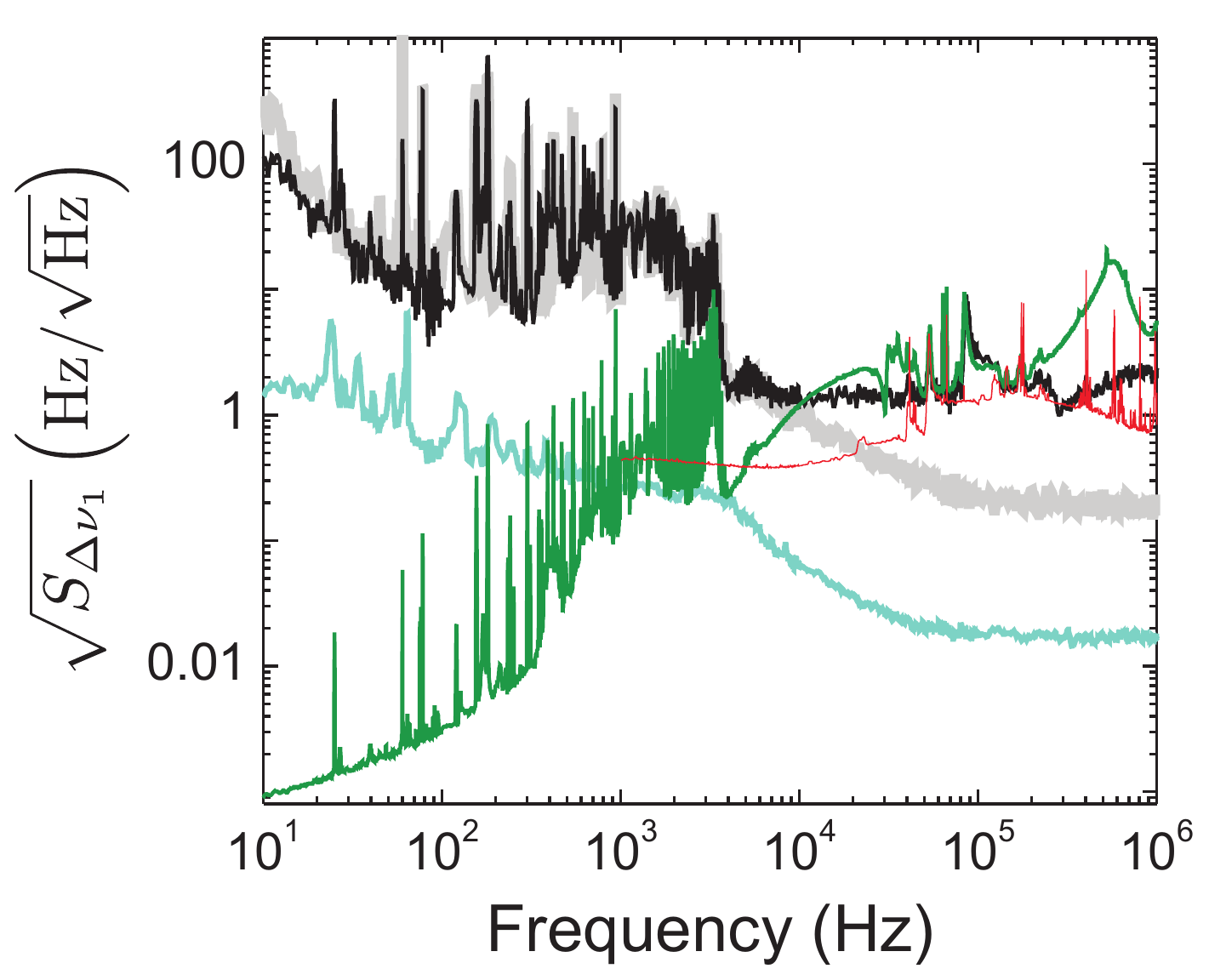}
\caption{Spectrum of line spacing fluctuations $S_{\Delta\nu_1}$.  The black and green lines show the free-running and stabilized line-spacing spectral density, respectively.  The broad resonance at 600 kHz in the green line coincides with a mechanical resonance of the fused quartz rod, which we speculate is weakly excited via the PZT.  The red line indicates the contribution from reference $f_1$.  The gray and blue lines show the predicted contributions from pump frequency and intensity noise, respectively.
\label{fig5} \vspace{-12pt}}
\end{figure}

To understand the pathway for future improvements in line spacing stability, we characterize the free-running noise spectrum of $\Delta\nu_1$; see the black curve in Fig. \ref{fig5}.  Our servo electronics reduce the frequency noise spectrum by up to $10^5$ within the 25 kHz bandwidth permitted by the PZT, and the spectrum after stabilization is shown by the green curve in Fig. \ref{fig5}.  Achieving further reduction in $S_{\Delta\nu_1}$ will depend on improvements among the feedback mechanism and the underlying source of the noise.  Here we focus on the latter.  In our current system, the primary contribution to $S_{\Delta\nu_1}$ is pump-frequency noise that maps onto the line spacing via a mostly constant relationship $\gamma_f = 10$ Hz$_{\Delta\nu_1}$/kHz.  This calibration was performed by modulating the pump frequency and recording the associated modulation in $\Delta\nu_1$.  By measuring the spectral density of pump-frequency noise and scaling it by $\gamma_f$, we obtain the gray curve in Fig. \ref{fig5}.  We also characterized the degree that pump intensity noise contributes to $S_{\Delta\nu_1}$.  In this case the mapping relationship is $\gamma_P=2$ kHz/mW, and it leads to the blue curve in Fig. \ref{fig5}.  It's surprising that intensity noise does not contribute more significantly to $S_{\Delta\nu_1}$, especially in light of previous data \cite{Papp2011}.  Still, our characterization of $S_{\Delta\nu_1}$ suggests a lower noise pump laser should be used in future experiments.  In particular with a factor of 10 improvement,
a residual frequency noise of $<100$ $\mu$Hz/$\sqrt{\rm{Hz}}$ at a 10 Hz offset from the 32.6 GHz line-spacing signal would be possible.  Access to microwave signals with such high spectral purity would enable interesting scientific and technological applications \cite{Fortier2011}.  Noise contributions from our measurement system also appear in $S_{\Delta\nu_1}$.  The red line shows the spectrum of $f_1$, which is generated by a high-performance commercial synthesizer.  This highlights the promise of microcomb technology, which here we demonstrate is already capable of producing signals commensurate with those of widely-used microwave signal generators.  
  
In conclusion, we have introduced new techniques for fabricating microresonators with $Q\gtrsim5\times10^8$, and for controlling the line spacing of parametric frequency combs created with them.  These resonators are exceptionally simple to create, and we have presented a deterministic procedure for varying their diameter.  Furthermore, we have reported a detailed study of microcomb line-spacing stabilization using piezoelectric mechanical control.  The achieved levels of absolute and residual fluctuations are respectively factors of 10 and 200 beyond all previous results \cite{DelHaye2008}.  This type of mechanical line-spacing control can easily be introduced into a variety of microcomb generators based on, for example, crystalline resonators \cite{Savchenkov2008} or integrated silicon nitride devices \cite{Okawachi2011}.  Our work has demonstrated microcomb residual noise that is capable of supporting modern frequency references beyond the $10^{-13}$ at 1 s level associated with traditional microwave oscillator technology.  Future work will focus on increasing the frequency span of the comb.

We thank Chris Oates and Gabe Ycas for their comments on this manuscript.  This work is supported by the DARPA QuASAR program and NIST.  This paper is a contribution of NIST and is not subject to copyright in the United States.  SP acknowledges support from the National Research Council.

\bibliographystyle{/Volumes/work1/bib_files/prsty}
\bibliography{/Volumes/work1/bib_files/sp_TF,/Volumes/work1/bib_files/sp_QOpt,/Volumes/work1/bib_files/sp_BEC}


\end{document}